\documentclass[journal]{vgtc}                     
\title{MusicJam: Visualizing Music Insights via \\ Generated Narrative Illustrations}

\author{Chuer Chen, Nan Cao, Jiani Hou, Yi Guo, Yulei Zhang, Yang Shi}

\authorfooter{
  \item
    Chuer Chen, Nan Cao, Jiani Hou, Yi Guo, Yulei Zhang, Yang Shi are with Intelligent Big Data Visualization Lab at Tongji University.\\E-mails:\{chuerchen1998, nan.cao\}@gmail.com, \{2231959, 2010937, zhangyulei\}@tongji.edu.cn, shiyang1230@gmail.com. Nan Cao is the corresponding author.
}

\abstract{%
Visualizing the insights of the invisible music is able to bring listeners an enjoyable and immersive listening experience, and therefore has attracted much attention in the field of information visualization. Over the past decades, various music visualization techniques have been introduced. However, most of them are manually designed by following the visual encoding rules, thus shown in form of a graphical visual representation whose visual encoding schema is usually taking effort to understand. Recently, some researchers use figures or illustrations to represent music moods, lyrics, and musical features, which are more intuitive and attractive. However, in these techniques, the figures are usually pre-selected or statically generated, so they cannot precisely convey insights of different pieces of music. To address this issue, in this paper, we introduce \name, a music visualization system that is able to generate narrative illustrations to represent the insight of the input music. The system leverages a novel generation model designed based on GPT-2 to generate meaningful lyrics given the input music and then employs the stable diffusion model to transform the lyrics into coherent illustrations. Finally, the generated results are synchronized and rendered as an MP4 video accompanied by the input music. We evaluated the proposed lyric generation model by comparing it to the baseline models and conducted a user study to estimate the quality of the generated illustrations and the final music videos. The results showed the power of our technique. 
}

\teaser{
  \centering
  \includegraphics[width=\linewidth]{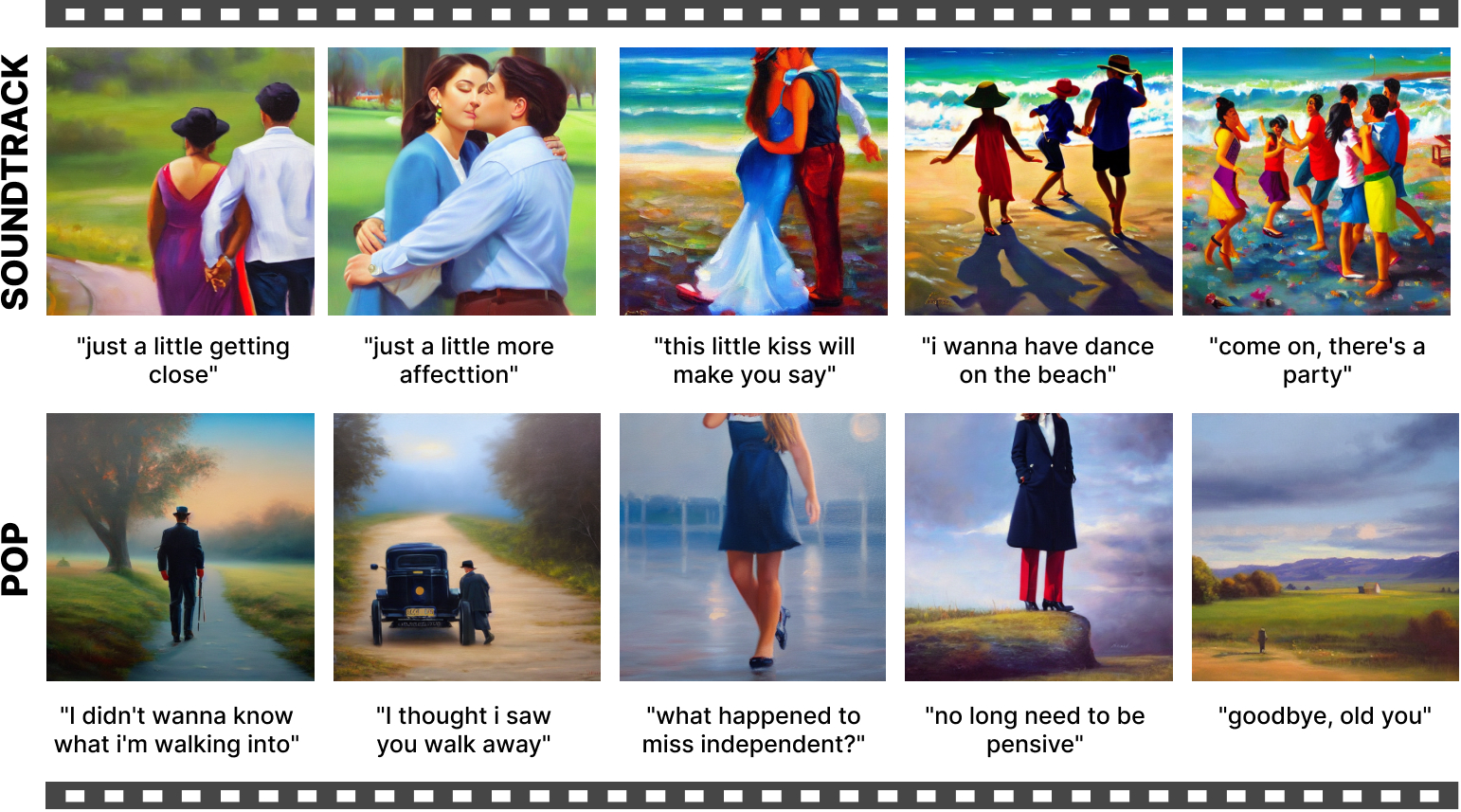}
  \caption{%
Music visualizations of soundtrack music and pop music with key illustrations and lyrics. The generated lyrics of the soundtrack music express a passionate feeling and the illustrations are oil painting style with vivid colors, which fit perfectly with the penetrating whistling music. The generated lyrics of the pop music describe a strong woman's story of moving on. The illustrations are oil painting style with neutral colors. The determined female depicted in the illustrations is quite in line with the tranquil music. The demos can be found in the following links: \url{https://www.youtube.com/watch?v=qC7_O8DzhyY} and \url{https://www.youtube.com/shorts/Gh9yXcEPrGQ}.%
  }
  \vspace{0.3em}
  \label{fig:teaser}
}

\graphicspath{{figs/}{figures/}{pictures/}{images/}{./}} 

\usepackage{tabu}                      
\usepackage{booktabs}                  
\usepackage{lipsum}                    
\usepackage{mwe}                       
\usepackage{amsmath}

\usepackage{mathptmx}                  
\usepackage{amssymb}

\newcommand{\etal}{{\it et~al.}\xspace}

\newcommand{\name}{MusicJam\xspace}

\begin{document}


\firstsection{Introduction}
\label{sec:intro}
\maketitle
Music visualization has long been an interesting topic in the field of information visualization~\cite{lima2021survey,khulusi2020survey}. Making the invisible music insights visible will inspire imagination, thus leading to a more immersive listening experience. Showing a coherent dynamic graphical representation to help visually narrate the insight of a piece of music while playing it has always been a great attraction for the audience. However, it is not an easy task, one may need to consider many aspects such as the content, melody, and genre so that a proper visual representation could be designed to catch the music spirit. 

During the past decades, many visualizations have been designed to help intuitively represent musical data~\cite{lima2021survey,khulusi2020survey} to aid learning~\cite{de2017understanding, benevento2020human, malandrino2018visualization, miller2018analyzing} or to provide a more immersive listening experience~\cite{hiraga2002performance, taylor2006real, graf2021audio, tatar2018revive}. Most of the existing visualizations are manually designed by following the visual encoding rules introduced in the field of information visualization, resulting in graphical visual representations that usually take effort to understand. Some researchers use figures or illustrations to represent music moods~\cite{cai2007automated, chen2008emotion, lehtiniemi2012using, lima2019visualizing}, lyrics~\cite{shamma2005musicstory, cai2007automated, xu2008automatic, funasawa2010automated, machida2011lyricon}, and musical features~\cite{cai2007automated, machida2011lyricon}, which are more intuitive and attractive. However, in these techniques, the figures are pre-selected and statically shared among all the pieces, so they cannot precisely convey the music insights. A technique that is able to dynamically generate consecutive and coherent illustrations for expressing and narrating the moods and insights of different music pieces is desired.

Generating narrative illustrations to represent the music insights is not an easy task. Three major challenges exist: (1) achieving high-quality cross-modal transformation from music signals to illustrations is still a technical challenge in the field, which has not been fully addressed; (2) generating narrative illustrations to interpret and reveal the music insights is not easy as it requires the generation of meaningful, consecutive, and coherent illustrations, which is a challenge for generation algorithm; (3) rendering the generated illustrations smoothly while playing the music need to synchronize the visuals with the audio tracks, which is not easy given a piece of pure music without any annotation.

To address the above challenges, we introduce \name, a system designed and developed for visualizing music via narrative illustrations that are automatically generated by analyzing the insight of the input music. To create meaningful and narrative visual representations, our system introduces a two-step illustration generation process: we first train a variational autoencoder~\cite{kingma2013auto} based on GPT2~\cite{radford2019language} to automatically generate meaningful lyrics for the input music and then generate a series of consecutive and coherent illustrations based on these lyrics by leveraging the pre-trained stable diffusion model~\cite{rombach2022sd}. The generation results of the lyrics and narrative illustrations are respectively evaluated via quantitative experiments and a user study. The contributions of the paper are as follows:

\begin{itemize}[leftmargin=10pt,topsep=1pt,itemsep=1px]
\item{\bf Lyric Generation Model.} We introduce a novel generative model, i.e., a variational autoencoder built on top of GPT-2, for generating meaningful and narrative lyrics given the input music. The generated lyrics are later used as part of prompts for generating illustrations based on the stable diffusion model.

\item{\bf Music Visualization System.} We developed an interactive system that integrates the above generative model to visualize music via consecutive narrative illustrations. Through the system, users are able to upload different music, set the style of illustrations, select the generated illustrations for different sections of the music, and adjust their order to compose a dynamic visualization that is synchronized with the music for producing a music video.


\item{\bf Evaluation.} We evaluate the quality (e.g., coherence and novelty) of the generated lyrics via a number of quantities metrics and estimate the quality of the generated illustrations via a user study comparing to a state-of-the-art music visualization tool, WZRD~\cite{wzrd}. The results verified the effectiveness of our technique.
\end{itemize}


\section{Related Work}
Our work draws inspiration from three areas, including music visualization, natural language generation, and text-to-image generation.

\label{sec:related}

\subsection{Music Visualization}
Music visualization (MusicVis) is an interdisciplinary field at the intersection of Information Visualization (InfoVis) and Music Information Research (MIR) ~\cite{lima2021survey}. Hiraga~\etal~\cite{1176199} defined MusicVis as the visual representation of a musical performance on a static or dynamic canvas expressively with computer graphics. Similar to the contribution of visualization in other fields, visualization techniques have the potential to enhance the understanding of musical data~\cite{munzner2014visualization} and enable individuals to identify underlying patterns in musical compositions~\cite{cooper2006visualization}.  


Lima~\etal~\cite{lima2021survey} provided a comprehensive survey of music visualization techniques and categorized them based on their visual representation forms, including colors, particles, shapes, line graphs, musical scores, glyphs, and pictures. In particular, colors are commonly used to represent features such as notes, pitch~\cite{lima2021survey}, volume~\cite{kosugi2010misual}, frequency power spectrum~\cite{ohmi2007music}, harmonic structure~\cite{7272579}, and mood of different passages~\cite{8812032}. Particles are utilized to depict dynamic changes in musical passages~\cite{lima2021survey,fonteles2013creating}. Shapes such as geometry and sine waves are frequently employed to illustrate features such as notes~\cite{cantareira2016moshviz}, pitch, volume~\cite{lima2021survey}, and rhythm~\cite{8812032}. Line graphs are used to represent features such as melody~\cite{8706365} and pitch~\cite{8336098} in music. Musical scores are used as a layout to analyze the music performance~\cite{kosugi2010misual}. Prisco~\etal~\cite{de2017understanding} utilized shapes and lines to visualize melody and musical intervals on sheet music, while encoding similar structures via~color~matrices. 

In various contexts, pictures are often utilized to portray the emotional ambiance of the music. For instance, Lehtiniemi~\etal~\cite{lehtiniemi2012using} employed image induction techniques to highlight the emotional aspects conveyed through musical performances. Similarly, Chin-Han Chen~\etal~\cite{chen2008emotion} attempted to acquire natural images for the purpose of detecting emotions. By aligning such images with music features that convey similar emotional characteristics, a sequence of visualizations in the form of slideshows can be generated. Additionally, the types of instruments played during a musical performance can also be depicted through pictures as a means of music visualization~\cite{969716}.


Our work is motivated by utilizing the pictures as the visual representing forms of music visualization to enhance the listening experience of music performances through the perceptual phenomenon of synesthesia~\cite{cytowic2002synesthesia}. In contrast to existing works~\cite{lehtiniemi2012using, chen2008emotion}, our technique automatically generates consecutive illustrations based on the input music, which is more flexible and meaningful.  

\subsection{Natural Language Generation}
Natural Language Generation (NLG) is defined as the task of generating text based on various forms of inputs~\cite{gatt2018survey}. It has been a popular academic trend for several decades with the development of successful deep-learning techniques. In general, NLG can be broadly divided into two categories: text-to-text generation~\cite{raffel2020exploring, xue2020mt5, cao2020factual, lewis2019bart,yang2018unsupervised, fu2018style, chen2018adversarial,li2013story, fan2018hierarchical} and data-to-text generation~\cite{lu2017exploring, lu2017knowing, anderson2018bottom, wang2018reconstruction, pei2019memory, sun2019videobert,hu2020makes, su2021bert}. 
Text-to-text generation aims to extract information from the input text and generate new text, such as text summarization~\cite{cai2019improving, cao2020factual, lewis2019bart}, text style transfer~\cite{yang2018unsupervised, fu2018style, chen2018adversarial}, and story generation~\cite{li2013story, fan2018hierarchical}. Conversely, data-to-text generation focuses on comprehending non-linguistic information within the inputs and generating the corresponding text, including the sub-tasks such as image caption~\cite{lu2017exploring, lu2017knowing, anderson2018bottom}, video caption~\cite{wang2018reconstruction, pei2019memory, sun2019videobert}.

Lyrics generation is a challenging task aimed at authoring lyrics based on music-related inputs~\cite{potash2015ghostwriter,wu2019hierarchical,wang2019theme,fan2019hierarchical, chen2020melody, vechtomova2021lyricjam, ma2021ai}. 
For instance, Wu~\etal~\cite{wu2019hierarchical} utilized an RNN-based Seq2Seq model to generate the next lyric lines with the previous lyric as input. To further improve the generation quality, several studies~\cite{chen2020melody, ma2021ai, vechtomova2021lyricjam} have introduced models that are conditioned on melody to better capture the correlation between lyrics and the melody. For instance, Chen~\etal~\cite{chen2020melody} trains an end-to-end SeqGAN model conditioned on melody represented in the form of numbered musical notation. Ma~\etal~\cite{ma2021ai} propose AI-Lyricist to generate lyrics given a required vocabulary and a MIDI file as inputs. It consists of a music structure analyzer, a music-lyrics embedding model, and a SeqGAN-based lyrics generator. LyricJam~\cite{vechtomova2021lyricjam} receives a live melody from a jam session and generates lyrics lines that are congruent with emotions and moods evoked by~the~music.

To the best of our knowledge, few studies have investigated how to generate coherent lyrics from musical input, which is a music-to-text generation task. To fill this gap, we proposed a variational autoencoder model with a cross-attention mechanism to realize multi-modal lyrics generation from input music.

\subsection{Text-to-Image Generation}
Text-to-Image Generation refers to the process of generating images based on textual descriptions. In recent years, various text-to-image models have been developed~\cite{tao2020dfgan, radford2021clip,xu2018attngan,ding2021cogview,rombach2022sd,saharia2022imagen,ramesh2022dalle2} and have gained widespread use in the creative community. For instance, CLIP~\cite{radford2021clip} is a pre-training model that has learned the correlation between text and images from a dataset comprising 400 million pairs of text and images. DALL-E~\cite{ramesh2021dalle} involves a two-stage process of compressing images into tokens using a dVA~\cite{dVAE}, and then using a transformer to simulate the joint distribution of text and image tokens. Additionally, diffusion-based generative models have shown promise in generating realistic images from natural language prompts. These models first disassemble the structure of the data distribution and then gradually restore it. DALL-E2~\cite{ramesh2022dalle2} is an example of a model that employs a diffusion model as a decoder to generate higher-quality images. Similarly, Imagen~\cite{saharia2022imagen} employs a frozen text encoder to embed the input text and uses a conditional diffusion model to map the text embedding to~an~image.

In this study, we utilized Stable Diffusion~\cite{rombach2022sd} to generate illustrations from the lyrical content. This method implements a diffusion model within the latent space of a pre-trained auto-encoder, and leverages a variational auto-encoder to decode the representation into the final image. Stable diffusion has demonstrated superior performance in image generation tasks compared to other current methods and is capable of generating high-quality images within 10 seconds, which meets the demands of~our~research.


\section{System Design}
\label{sec:system}

In this section, we describe the design requirements of the \name system, followed by an introduction to the system's architecture and user interface.

\begin{figure}[tb]
  \centering
  \includegraphics[width=\linewidth]{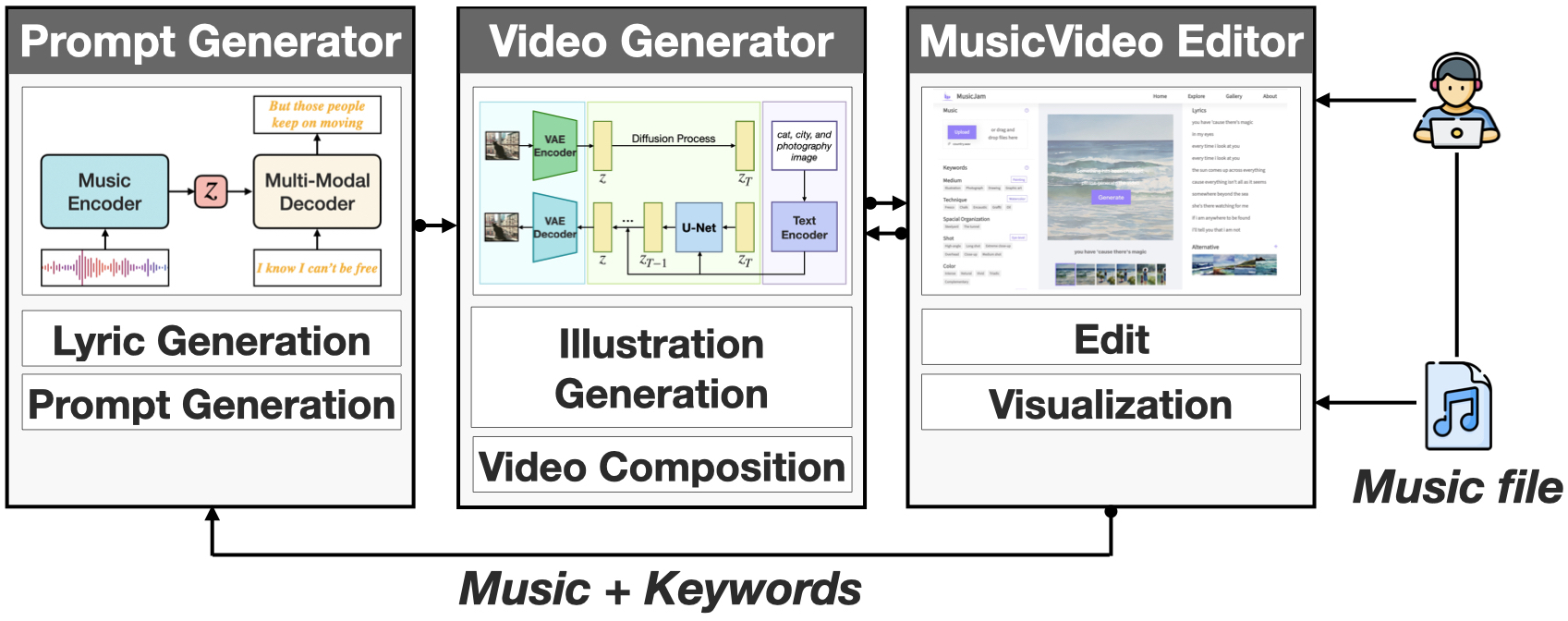}
  \vspace{-1.5em}
  \caption{The architecture and running pipeline of the \name system.} 
  \label{fig:architecture}
  \vspace{-1.5em}
\end{figure}

\subsection{Design Requirements}
\label{sec:designrequirements}
The proposed \name system has been designed to fulfill several fundamental design requirements for generating a high-quality music visualization, which are summarized as follows:

\begin{enumerate}[topsep=0pt,itemsep=2px,partopsep=1pt,parsep=3pt]

\item[{\bf R1}] {\bf Representing the music via meaningful illustrations.} The system should be able to represent the music in meaningful and inspiring visual illustrations that are able to reveal the moods and narrate the insights of the music.

\item[{\bf R2}] {\bf Coherent dynamic visual representations.} Besides being meaningful, the system should be able to generate a sequence of coherent illustrations that change smoothly while playing the music to provide a better listening experience.

\item[{\bf R3}] {\bf Flexible editing on the generation results.} The system should be flexible enough to let the users modify the generated results based on their own preferences to produce a high-quality~music~video.
\end{enumerate}


To fulfill the design requirements, the design of \name system consists of three major modules (Fig.~\ref{fig:architecture}): (1) the prompt generator, (2) the music video generator, and (3) the music video editor. In particular, a user first uploads a music file and selects a set of initial keywords to determine the visual style of the generated illustrations. The uploaded music file will be used by the Prompt Generator, which automatically generates lyrics as a narrative of the music content. These lyrics are further used to guide the generation of visual illustrations so that the music content could be meaningfully represented in the form of visual narrations (\textbf{R1}). In particular, these lyrics are stitched with keywords to form lyrics-to-illustration prompts, which is used by the Video Generator to create narrative illustrations based on the stable diffusion model and then synthesize the illustrations and music into a music video (\textbf{R2}). Finally, the user can play the generated music video in the Music Video Editor or make modifications by reordering or substituting illustrations (\textbf{R3}). 


\subsection{User Interface and Interactions}
\begin{figure}[!tbh]
  \centering
  \includegraphics[width=\linewidth]{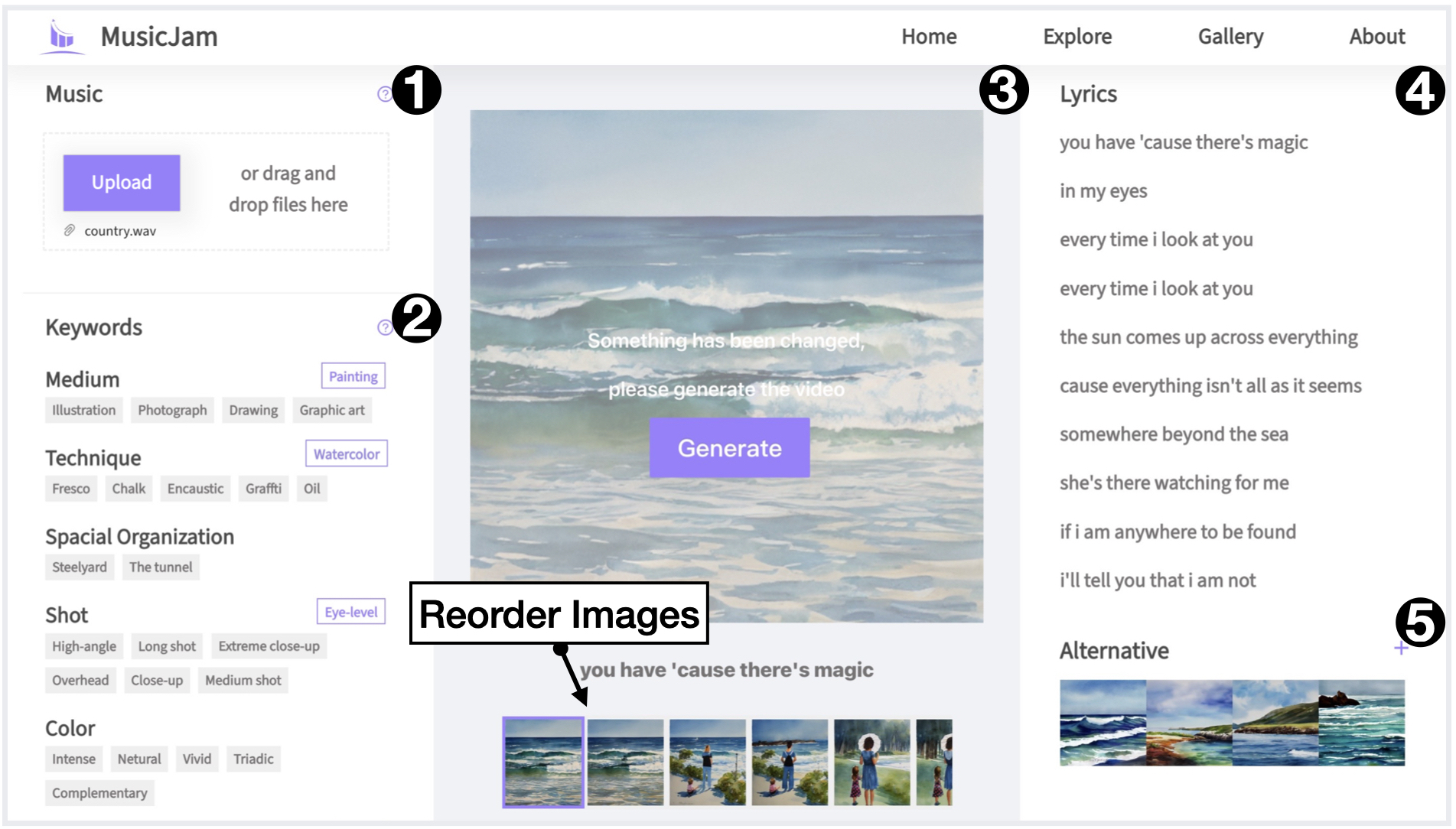}
  \caption{\name system consists of five views: (1) the music view for uploading music files, (2) the keywords view for selecting keywords, (3) the visualization view for playing and editing the music video, (4) the lyrics view for displaying generated lyrics and (5) the alternative view for adjusting illustrations. } 
  \label{fig:interface}
  \vspace{-1.5em}
\end{figure}

In this section, we introduce the design of the music video editor, which provides an intuitive experience for users to generate, watch, and refine the music visualization as depicted in Fig~\ref{fig:interface}. 


\label{sec:interface}

To generate a music visualization, users need to first upload a music file in MP3 or WAV format either by choosing the file from a folder or directly dragging the file into the uploading box (Fig.~\ref{fig:interface}(1)). Once successfully upload, users can select their preferred keywords to control the styles of the illustration to be generated from a list presented in the keywords view (as depicted in Fig.~\ref{fig:interface}(2)). The system offers six categories of style keywords, including media, technique, color, spatial organization, and light, which have a substantial impact on the quality and style of the generated illustrations. The users are allowed to interactively adjust the keywords so that the generated visualization could better align with their desired outcomes. 

Once the necessary configurations have been made, by clicking the ``Generate" button (Fig.~\ref{fig:interface}(3)), the system will proceed to generate the music video on the server and provide the results in a matter of moments. The generated music video is displayed in the playback window with the corresponding lyrics displayed as subtitles at the bottom, which scroll in synchrony with the video. These lyrics are also presented in the lyrics view (Fig.~\ref{fig:interface}(4)) on the right side of the interface, which narrates the insight of the music as a whole.

If users are not satisfied with the generated visualization result, they can reorder the illustrations by dragging them in the list at the bottom of the visualization view. Additionally, users can select an unsatisfied illustration and replace it by using the substitutes from the alternative view (Fig.~\ref{fig:interface}(5)). Upon completion of the editing processes, the users can click the ``Generate" button to obtain the modified visualization.

\section{Lyric Generation Model}
In the system, we introduce a novel variational autoencoder that is built upon GPT-2~\cite{radford2018improving}  for generating lyrics that interpret and narrate the input music by balancing between coherence and diversity. 
In this section, we first briefly introduce the variational autoencoder. Then, we interpret the design of our model, including the music encoder, the multi-modal decoder, and the~training~objective.


\label{sec:model}

\subsection{Variational Auto-Encoder}
\begin{figure}[!tbh]
  \centering
  \includegraphics[width=\linewidth]{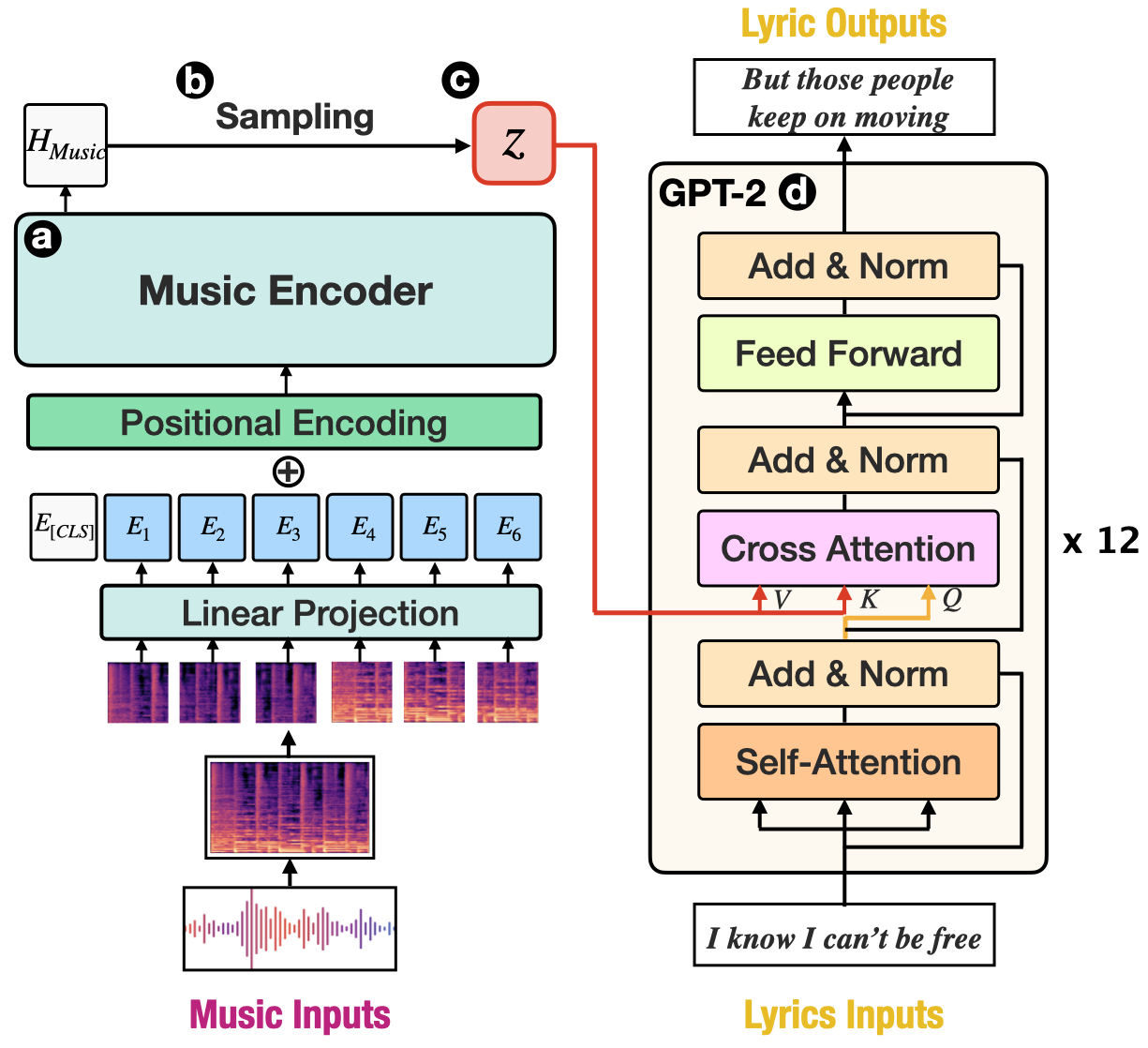}
  \caption{The architecture of the lyric generation model.} 
  \label{fig:model}
  \vspace{-1.5em}
\end{figure}

The variational autoencoder~(VAE)~\cite{kingma2013auto} is a prominent framework utilized in various generation tasks. The underlying principle of this framework involves encoding the input data $x$ into a latent distribution $z$ that approximates the prior distribution $p(z)$, followed by decoding the latent distribution to generate a sample of $x$. The objective function of VAE, called evidence lower bound (ELBO), consists of the reconstruction loss that measures the dissimilarity between the actual output and the target output, and the Kullback-Leibler (KL) divergence that assesses the difference between the $p(z)$ and the distribution produced by the encoder. The ELBO is represented as:

\begin{equation}
    ELBO = E_{{q_\phi}(z|x)}[\log{{p_\theta}(x|z)}]-KL({q_\phi}(z|x)||p(z))
\label{eq:reward}
\end{equation}
where $p(z)$ is the prior distribution which is commonly assumed as standard Gaussian; ${p_\theta}(x|z)$ is the decoder to reconstruct $x$ from the latent variable $z$; $\phi$ and $\theta$ represent the parameters of encoder and decoder respectively.

An essential aspect of VAE involves learning a posterior distribution $q_\phi(z|x)$, which enables the model to approximate the true prior distribution~$p(z)$~effectively. Given that benefit, we introduce a novel variational autoencoder that is built upon GPT-2~\cite{radford2018improving} to facilitate the transformation of music to lyrics. The model, as depicted in Figure~\ref{fig:model}, involves a \textbf{music encoder}~(Fig.~\ref{fig:model}(a)) to learn a latent vector $z$~(Fig.~\ref{fig:model}(c)) from a music clip, which is then utilized to guide the lyric generation through a \textbf{multi-modal decoder}~(Fig.~\ref{fig:model}(d)). This process incorporates a sampling of $z$, as shown in Fig.~\ref{fig:model}(b), which introduces an element of randomness to the music-to-lyric conversion, thereby ensuring that the generated lyrics exhibit diversity given~the~same~input.

\subsection{Music Encoder}

To ensure that the generated lyrics are aligned with the accompanying music, we have integrated a music encoder into our model, as illustrated in Figure~\ref{fig:model}(a). The purpose of this encoder is to extract the hidden representation from the music, denoted by $H_{\text{music}}$, which is subsequently used to guide the process of lyric generation. For the implementation of the music encoder, we have employed the Audio Spectrogram Transformer~(AST)~\cite{gong2021ast}, an attention-based model that can capture global features from an audio spectrogram.

As depicted in Figure~\ref{fig:model}, given a music clip $m_x$, we first transform the clip into a mel spectrogram and subsequently divide it into a sequence of overlapping patches. After that, to ensure that the information is captured accurately, we apply a linear projection layer to each patch, thereby generating a corresponding patch embedding vector. Then, we incorporate positional information by adding a positional embedding to each patch. Moreover, we prepend a [CLS] token to the sequence, enabling us to obtain a global representation of the sequence from the encoder. Finally, the sequence is fed into the AST to generate $H_{\text{music}}$: 
 \begin{equation}
H_{\text{music}} = \text{encode}~(m_x)
\end{equation}
where $H_{\text{music}}$ refers to the hidden state of the [CLS] token obtained from the last~encoder~layer. Then, $H_{\text{music}}$ is transformed into two vectors, $\mu$ and $\sigma$, which are the parameters (i.e., mean values and standard deviations) of a set of normal distributions used to capture the distributions of music:
 \begin{equation}
    {\mu} = W_{\mu}H_{\text{music}}
\label{eq:reward}
\end{equation}
 \begin{equation}
    {\sigma} = \text{exp}(\frac{{W_{\sigma}H_{\text{music}}}}{2})
\label{eq:reward}
\end{equation}
where ${W_\mu}, {W_\sigma}$ are trainable weight matrices. Finally, a latent vector $z$ is randomly sampled from the distributions to generate the next lyric:
 \begin{equation}
    z = \mu + {\sigma}\odot{\epsilon}, ~\epsilon\sim{N(0,I)}
\label{eq:reward}
\end{equation}
where $\epsilon$ is a random vector sampled from the distribution $N(0, I)$ that ensures that $z$ is nondeterministic.

\subsection{Multi-Modal Decoder}
With the latent vector $z$ from the music encoder, we utilize a multi-modal decoder (as illustrated in Figure~\ref{fig:model}(d)) to generate lyrics conditioned on the musical input and previous lyrics. Specifically, we initialize the multi-modal decoder with GPT-2~\cite{radford2018improving}, a pre-trained transformer-based language model capable of generating coherent text based on prompts. The decoder accepts the latent vector $z$ and the previous lyric $l$ as inputs, and generates subsequent lyrics.

As depicted in Figure~\ref{fig:model}(d), the past lyric $l_\text{past}$ is regarded as a context and fed into the decoder to ensure coherence with the newly generated lyrics. The embeddings of $l_\text{past}$ first pass through a multi-head self-attention layer and a feed-forward neural network to have a hidden representation $h_\text{lyrics}$ that captures the global information. To ensure the generated lyrics correspond with the input music, we implement a cross-attention mechanism that links $h_\text{lyrics}$ and $z$, resulting in an enhanced hidden representation $h_\text{lyrics+music}$:
\begin{equation} 
h_\text{lyrics+music} = MultiHeadAttn(h_\text{lyrics},~ z,~z,~n)
\end{equation}
where $n$ is the number of attention heads. $h_\text{lyrics+music}$ is further calculated by a feed-forward neural network and a softmax layer to compute the output probabilities of the tokens in the vocabulary. In each round, the token in the vocabulary having the highest probability is chosen as the output of the model. In this way, the decoder generates a lyric line token by token in the following form:
\begin{equation}
\label{eq:decompose}
[\mathtt{t}_{1},\ldots, \mathtt{t}_{n}] \leftarrow \text{decode} (l_\text{past} ,z)
\end{equation}
where $\mathtt{t}_{i}$ indicates the $i$-th word of the generated lyric.




\subsection{Training Objective}
Similar to VAE, the training objective of our model consists of a reconstruction loss and a KL divergence, which is defined~as~follows:

 \begin{equation}
    {L_\theta}(x,y,z,\hat{y}) = {L_{recons}}(y,\hat{y})+\beta{KL({q_\phi}(z|x)||p(z))}
\label{eq:reward}
\end{equation}
where ${L_\text{recons}}(y,\hat{y})$ calculates the dissimilarity between the generated lyrics $y$ 
and the golden lyrics $\hat{y}$ annotated in our dataset; $KL({q_\phi}(z|x)||p(z))$ assesses the difference between the $p(z)$ and the distribution produced by the encoder; $\beta$ is a hyper-parameter controlling the loss contribution from the KL divergence.

\subsection{Dataset}
\label{sec:dataset}
To train our model, we collect a music-lyric dataset from DALI~\cite{meseguer2019dali}. DALI is an extensive collection of synchronized audio, lyrics, and musical notations that serves as a benchmark for the singing voice research community. It comprises 5358 songs featuring time-aligned vocal melody notes and lyrics categorized into four distinct levels of granularity. From this dataset, we specifically choose 2590 English songs belonging to various genres, including but not limited to pop, alternative, and rock. 

We separated the accompaniment from the song and partitioned each accompaniment into a collection of 5-second segments accompanied by synchronized lyric lines, producing a total of 73,775 pairs of music and lyrics. To preserve contextual information, each pair is labeled with the previous lyric line, which serves as the context~(i.e., the textual input to the decoder) for the lyric generation process. We used a special token ``<START>'' for the previous lyric line of the first lyric line. The music-lyric pairs are divided into two distinct sets, a training set and a test set, in equal proportion according to genres. The training set comprises 2072 songs and 58452 music-lyric pairs, while the test set contains 518 songs and 15323 music-lyric pairs.

\subsection{Implementation}
In the proposed model, the music encoder and multi-modal decoder are equipped with 12 layers each and utilize multi-head attention layers composed of 12 heads. To tackle the potential issue of KL vanishing that may arise during the training phase, the KL term weight $\beta$ is maintained at 1e-5 during the first half of the training period, after which it is linearly increased to a value of 1 for the remaining duration of the training phase. Furthermore, to enhance the creativity of the lyric generation process, we utilize top-k ($k=100$) and top-p ($p=0.95$) sampling techniques combined with temperature smoothing~($T=0.95$) in the multi-modal decoder. 

Our model was implemented by PyTorch and trained on an Ubuntu server with an Nvidia Tesla-V100 16GB graphic card. The model was trained via 40 epochs with a batch size of 32. We used the Adam optimizer with a learning rate of 5e-5.

\section{Visualization}

\label{sec:vis}
We visualize the input music by generating narrative illustrations and then compose them with the music into a music video. To narrate the music, the generated illustrations must be meaningful and coherent, which can be achieved by using meaningful and coherent lyrics to guide the generation process. To this end, we employ stable diffusion~\cite{rombach2022sd}, the state-of-the-art text-to-image generation model, and carefully investigate how to use text prompts to precisely control the generation results. A series of experiments for prompt engineering was thus conducted. In this section, we first present the details of the experiments of prompt engineering. We then introduce the stable diffusion model followed by introducing the illustration interpolation method used to improve the coherence of the music video. Finally, we describe the method of generating a music video from the input music and generated~illustrations.


\subsection{Prompt Engineering}
\begin{table*}[!ht]
\centering
\label{tab:prompts}
    
    \setlength\aboverulesep{0pt}
    \setlength\belowrulesep{0pt}
    \def\arraystretch{1.5}

\begin{tabular}{@{}lccccc@{}}
\toprule
\multicolumn{1}{c}{\textbf{Medium}}                                           & \textbf{Technique}                                      & \textbf{Spatial Composition}      & \textbf{Shot}                           & \textbf{Color}                             & \textbf{Light}                         \\ \midrule
\multicolumn{1}{c}{{\color[HTML]{0E5ACA} Painting}}                           & {\color[HTML]{0E5ACA} Oil, Encaustic, Fresco,}          & {\color[HTML]{0E5ACA} Steelyard}  & {\color[HTML]{0E5ACA} Overhead shot}    & {\color[HTML]{0E5ACA} Complementary color} & {\color[HTML]{0E5ACA} Warm light}      \\
\multicolumn{1}{c}{{\color[HTML]{0E5ACA} Drawing}}                            & {\color[HTML]{0E5ACA} Graphite pencil, Wax color,}      & {\color[HTML]{0E5ACA} The tunnel} & {\color[HTML]{0E5ACA} High-angle shot}  & {\color[HTML]{0E5ACA} Tradic color}        & {\color[HTML]{0E5ACA} Day light}       \\
\multicolumn{1}{c}{{\color[HTML]{0E5ACA} Graphic art}}                        & {\color[HTML]{0E5ACA} Pastel, Digital tool, Ink,}       &                                   & {\color[HTML]{0E5ACA} Eye-level shot}   & {\color[HTML]{0E5ACA} Intense color}       & {\color[HTML]{0E5ACA} Moon light}      \\
\multicolumn{1}{c}{\color[HTML]{333333} Photograph} & {\color[HTML]{333333} Polaroid, Monochrome,}            &                                   & {\color[HTML]{333333} Close-up}         & {\color[HTML]{0E5ACA} Neutral color}       & {\color[HTML]{333333} Soft light}      \\
\multicolumn{1}{c}{{\color[HTML]{333333} Illustration}}                       & {\color[HTML]{333333} Long exposure, Color splash,}     &                                   & {\color[HTML]{333333} Extreme close-up} &                             {\color[HTML]{333333} Vivid color}       & {\color[HTML]{333333} Ambient light}   \\
                                                                              & {\color[HTML]{333333} Tilt-shift, Wide-angle, Vector,}  &                                   & {\color[HTML]{333333} Medium shot}      &                                            & {\color[HTML]{333333} Ring light}      \\
                                                                              & {\color[HTML]{333333} Telephoto, Bokeh, Caricature,}    &                                   & {\color[HTML]{333333} Long shot}        &                                            & {\color[HTML]{333333} Sun light}       \\
                                                                              & {\color[HTML]{333333} Children’s book, Comic, Chalk,}   &                                   &                                         &                                            & {\color[HTML]{333333} Cinematic light} \\
                                                                              & {\color[HTML]{333333} Propaganda poster, Movie poster,} &                                   &                                         &                                            &                                        \\
                                                                              & {\color[HTML]{333333} Water colors, Graffiti, Ukiyo-e,} &                                   &                                         &                                            &                                        \\
                                                                              & {\color[HTML]{333333} Psychedelic art, Splash art}      &                                   &                                         &                                            &                                        \\ \bottomrule
\end{tabular}
\caption{All the prompt keywords in MusicJam. The blue words are selected in Experiment 1, others are selected from prompt book~\cite{promptbook}.}
\label{tab:keywords}
\vspace{-1em}
\end{table*}                                          

To generate high-quality illustrations with the stable diffusion model, two experiments were conducted to explore the methodologies for selecting or constructing appropriate prompts. In particular, in addition to lyrics, it was desired to incorporate keywords into the prompts to manipulate the style, format, or perspective of the illustrations. To this end, two experiments were conducted. The first experiment aimed at identifying keywords that are suitable as prompts, as they have a discernible effect on the generated illustrations. The second experiment focused on determining the optimal way to arrange these keywords into a valid prompt.

\textbf{Experiment 1: Keywords Selection.} In order to select valid keywords, a set of candidate keywords was first collected. Subsequently, the impact of these keywords on the generation process was evaluated by two designers, and finally, keywords that were found to have a discernible impact on the generated illustrations were selected~as~prompts.

\underline{\textit{Collection.}} 
To achieve a more extensive collection of candidate keywords, the framework proposed by Rose~\etal~\cite{rose2022visual} was employed to guide the collection process. Specifically, the framework pays attention to the technological modality and compositional modality of images. The technological modality refers to the media~(e.g. painting, drawing) and techniques~(e.g. watercolor, oil in painting) used in the production of the image. The compositional modality mainly includes the crucial components of the image. Based on this framework, we summarize six important dimensions of images, namely medium, technique, spatial organization, shot, color, and light. With these dimensions, we collect a set of candidate keywords from related sources, including~\cite{promptbook}, \cite{rose2022visual}, \cite{edgar2005composition}, and \cite{gurney2010color}. Finally, 82 candidate keywords were gathered for~further~evaluation.


\underline{\textit{Procedure.}} In order to evaluate the impact of the candidate keywords on the generated illustrations, an experiment was carried out. This process involved using the candidate keywords to generate illustrations, which were then subjected to inspection by two designers. Specifically, 10 lyrics from the song \textit{``Love Story"} were selected and combined with keywords to form prompts. Here, we leave out 32 candidate keywords from prompt book~\cite{promptbook}~(listed in Table~\ref{tab:keywords}) as they have been proven to be effective for stable diffusion. To this end, 500 prompts were formulated by combining each of the 50 candidate keywords with 10 lyric lines. As a result, 493 illustrations were produced through the stable diffusion model and were subjected to further analysis. Some generated illustrations were excluded from the study because they were judged as ``Not Safe/Suitable For Work'' materials~\cite{nsfw}.

With the generated illustrations, we invited two designers with backgrounds in the art to assess the impact of the keywords. We first explained the meaning of candidate keywords and show how they represent in other images. Then, we displayed 493 illustrations to the participants one by one and asked them to annotate whether the generated illustration is noticeably related to its keyword.

\underline{\textit{Results.}}
Based on the annotations provided by the two designers, it was found that their assessments were relatively consistent, with 85.1\% of the annotations being in agreement. The annotations were further validated using Cohen's kappa~\cite{cohen1960coefficient}, which yielded a result of $\kappa$=0.66, indicating a level of agreement between the two~designers'~annotations.

Next, we selected keywords that had a noticeable impact on the generation based on the annotation. This was accomplished by counting the number of illustrations generated by each keyword that was recognized by both designers as being impacted. Only those keywords that were determined to have an impact on five or more illustrations were deemed to be valid. Finally, we retained 23 keywords~(listed in Table~\ref{tab:keywords}) as prompt keywords to control the illustration generation.

\textbf{Experiment 2: Keywords Permutation.} In this study, we aim to investigate the impact of the arrangement of keywords on the generation of illustrations. To accomplish this, we examine various permutations of keywords in order to determine how the modification of prompts controls the generation. 


\underline{\textit{Procedure.}} To examine the influence of the arrangement of keywords on illustration generation, we conducted an experiment that involved testing different permutations of keywords. The experiment was based on the results of the first experiment, in which keywords were selected that had the greatest impact on illustration generation. A designer was consulted to arrange the keywords based on their significance to the image, in the following order: painting, steelyard composition, eye-level shot, intense color, and warm light. In order to compare the impact of the positions, we placed the same keyword in the first and last position respectively to form two permutations as a comparison group, and composed five groups in total. The same ten lyrics used in the first experiment were combined with these permuted keywords to form 100 prompts, which were then used to generate 100 illustrations.

Two designers from the first experiment were invited to judge if the position of keywords affects the illustration. The participants were presented with pairs of illustrations generated from two prompts, each belonging to a comparison group. Then, we asked them to indicate which illustration was better than the other. In instances where no notable difference was perceived between the two illustrations, the participants were instructed to annotate ``none''.

\underline{\textit{Results.}} With the collected annotations, we binned the generated illustrations based on whether they were annotated as ``significantly better''. We denoted the "significantly better'' illustrations as prominence and others as normal. We observed that the participants had an agreement of 84.0\% on the assessments. We further reported a Cohen's kappa of 0.34, which represented fair agreement across the generations. Next, we built a contingency table based on the annotations and performed a Chi-square test on it. The test result ($\chi^2(1, N = 100)=2.747, p=0.432$) indicated that there was no significant difference between the quality of illustrations generated by the two permutations. Hence, we concluded that the position of the keyword had no significant effect on the illustration generation. Therefore, we did not take the position of keywords into consideration when constructing a prompt.

\subsection{Illustration Generation}
In \name, we utilize the stable diffusion model~\cite{rombach2022sd} to produce high-quality illustrations. This cutting-edge text-to-image technique is capable of generating realistic images based on textual prompts and has demonstrated superior performance in comparison to generative adversarial networks in image synthesis tasks. The stable diffusion model operates by gradually adding random noise to the data, and then learning to reverse the diffusion process on a compressed latent space to construct the desired data samples from~the~noise.

\begin{figure}[!tbh]
  \centering
  \includegraphics[width=\linewidth]{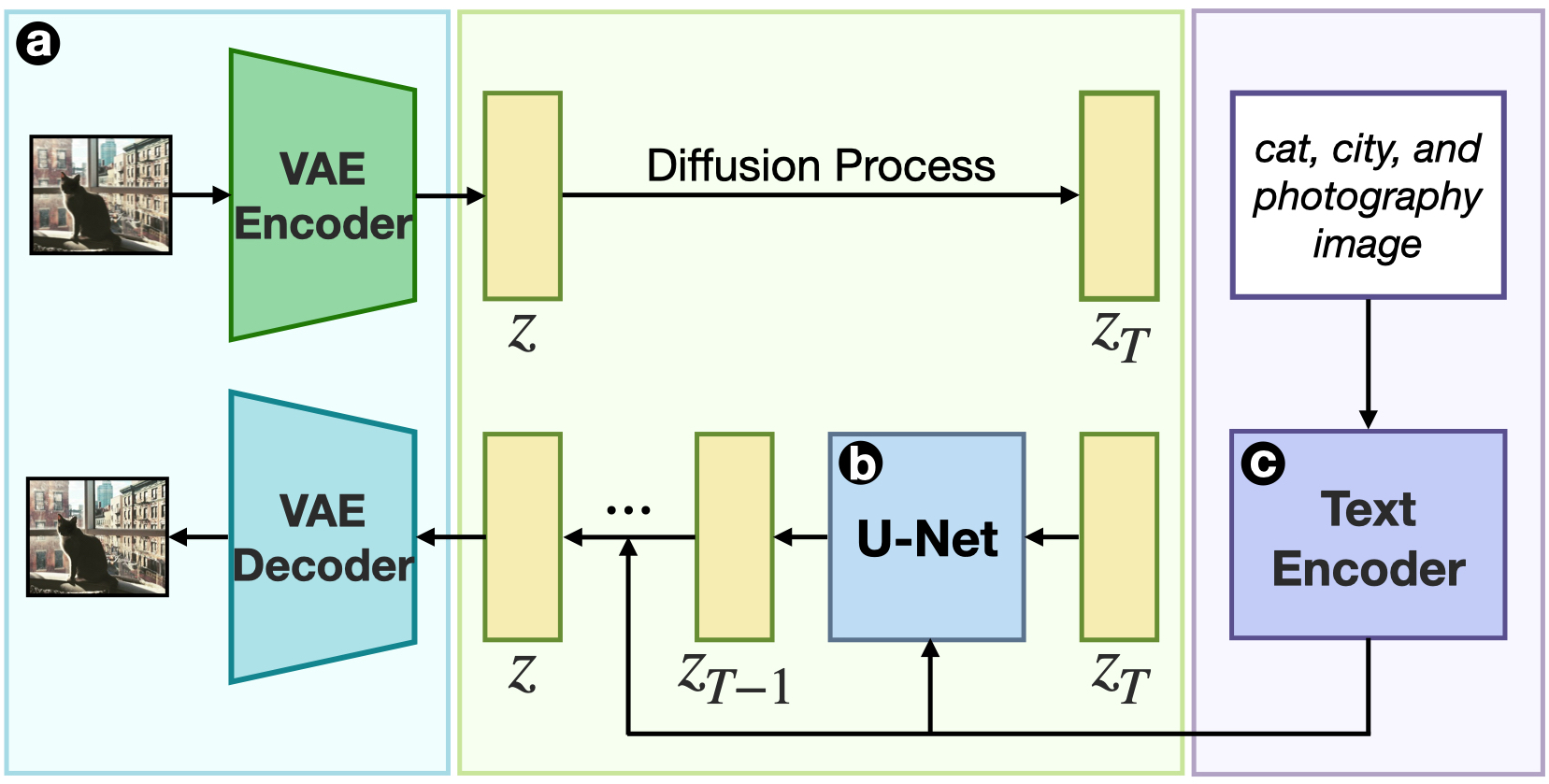}
  \caption{The architecture of the stable diffusion model.} 
  \label{fig:diffusion}
  \vspace{-1em}
\end{figure}

 The stable diffusion model consists of three main components: (1) the Variational Autoencoder (Fig.~\ref{fig:diffusion}(a)) for compressing the pixel space to a lower dimensional latent space; (2) the U-Net (Fig.~\ref{fig:diffusion}(b)) for denoising random Gaussian noise to construct the image representation; (3) the Text Encoder (Fig.~\ref{fig:diffusion}(c)) for transforming the input prompt to the embedding space. The denoising process is conditioned on the text embeddings via cross-attention layers in the U-Net so that the image generation can be guided by text prompts. Therefore, the key factor that determines the quality and the content of the generated illustrations is the appropriately described prompts.

\subsection{Illustration Interpolation}
With the stable diffusion model, \name is 
able to generate illustrations and integrate them into a music video. To make the video more coherent, we employ an illustration interpolation method specifically for stable diffusion to synthesize more illustrations between two adjacent lyric lines.  

The stable diffusion model employs a text encoder~(Fig.~\ref{fig:diffusion}(c)) to convert the prompts into embeddings, which are then input into the conditional denoising U-Net to regulate the illustration synthesis process. Therefore, interpolation between two prompts is facilitated through the use of intermediate text embeddings. To ensure a harmonious relationship between the video and music, the text embedding interpolation is calculated based on the beat of the music.  Specifically, the interpolation process consists of four steps. First, we extract the percussive wave from the input music and convert it to a mel spectrogram. Second, we obtain the maximum sound amplitude of the mel spectrogram at each time step and then normalize it. After that, we re-scale the normalized maximum sound amplitude to the number of interpolation steps $N$. Finally, we accumulate it and normalize it again to get the interpolation weights. We calculate the interpolated embedding vector as follows:
\begin{equation}
    {e_i} = \text{text-encoder}(p_i)
\label{eq:textembedding}
\end{equation}
 \begin{equation}
    {e_k} = {e_i}+{w_k}\cdot{(e_{i+1}-e_i)}
\label{eq:interpolation}
\end{equation}
where ${p_i}$ indicates the $i$-th prompt; ${e_i}$ is the embedding vector extracted from the text encoder of stable diffusion model (Fig.~\ref{fig:diffusion}(c)); $e_k$ ($k\in{[1,\cdots,N]}$) is the $k$-th interpolated embedding vector between ${e_i}$ and ${e_{i+1}}$; $w_k$ is the $k$-th interpolation weight. 

In addition to the prompt, the latent seed is another input of the stable diffusion model during inference. The latent seed is used to generate a random latent noise, which will be further denoised in the U-Net conditioned on text embedding. Therefore, we apply Spherical Linear Interpolation~\cite{shoemake1985animating} method to generate the interpolated latent noise:
\begin{equation}
    {\Omega} = \arccos{{n_i}\cdot{n_{i+1}}}
\label{eq:omega}
\end{equation}
 \begin{equation}
    {n_k} = {\frac{\sin{[1-w_k]\Omega}}{\sin{\Omega}}}{n_i} + {\frac{\sin{{w_k}\Omega}}{\sin{\Omega}}}{n_{i+1}} 
\label{eq:interpolation}
\end{equation}
where ${n_i}$ indicates the $i$-th latent noise; $\Omega$ is the angle between ${n_i}$ and ${n_{i+1}}$; $n_k$ ($k\in{[1,\cdots,N]}$) is the $k$-th interpolated latent noise between ${n_i}$ and ${n_{i+1}}$;

\begin{figure}[!tbh]
  \centering
  \includegraphics[width=\linewidth]{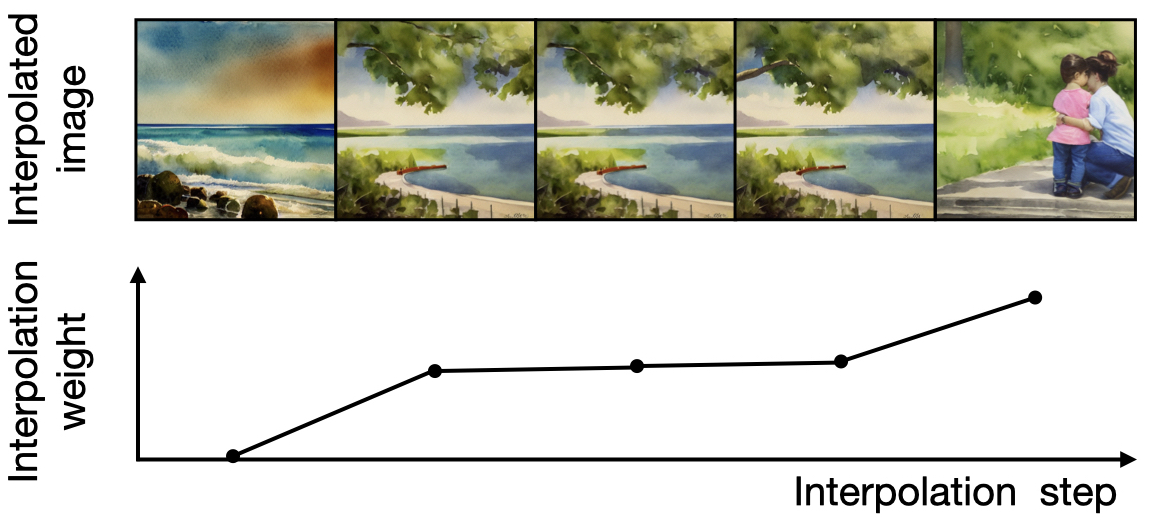}
  \vspace{-2em}
  \caption{The interpolation results in 5 seconds. ${p_i}=$"somewhere beyond the sea", ${p_{i+1}}=$"she's there watching for me"} 
  \label{fig:interpolation}
  \vspace{-1em}
\end{figure}

With the interpolated text embeddings and latent noises, we can generate interpolated illustrations by stable diffusion. We present an example of interpolation in Fig.~\ref{fig:interpolation}. When the amplitude of the music suddenly becomes larger, the interpolation weight is sharply increased and the content of the illustration changes a lot. In contrast, when the interpolation weights are similar, the interpolated illustrations look almost the same. With the interpolation method, the movement of the video is changing according to the music's beat, which provides a smooth and rhythmic visual effect for users.

\begin{table*}[!htb]
\centering

    \setlength\aboverulesep{0pt}
    \setlength\belowrulesep{0pt}
    \def\arraystretch{1.5}
    \begin{tabular}{ccccccccc}
    
\toprule
\textbf{Model}    & \textbf{BLEU-2}   & \textbf{BLEU-3}   & \textbf{Distinct-2} & \textbf{Distinct-3} & \textbf{Novelty-2} & \textbf{Novelty-3} &  \textbf{Coherence} & \textbf{CLIPScore} 
\\\midrule

$\mathit{B}_1$ & 0.323 & 0.191 & 0.120 & 0.241 & 0.347 & 0.563  & 0.126 & 22.62   \\
$\mathit{B}_2$ & 0.319 & 0.189 & 0.175 & 0.348 & 0.532 & 0.777  & 0.159 & 23.03                                      \\
\textbf{\textcolor{teal}{Lyric Generation Model}} & \textbf{\textcolor{teal}{0.325}} & \textbf{\textcolor{teal}{0.192}} & \textbf{\textcolor{teal}{0.215}} & \textbf{\textcolor{teal}{0.394}} & \textbf{\textcolor{teal}{0.618}} & \textbf{\textcolor{teal}{0.796}} & \textbf{\textcolor{teal}{0.173}} & \textbf{\textcolor{teal}{23.17}}

    \\\bottomrule    
    
\end{tabular}
\caption{Results of our proposed model compared to baselines on quantitative evaluations}
\label{tab:evaluation}
\vspace{-1em}

\end{table*}

\subsection{Music Video Generation}
The last step is to compose the music and the generated illustrations into a music video. Currently, the stable diffusion model takes about 8 seconds to generate an illustration. To minimize the user's waiting time, we control the number of generations and design to generate five illustrations with two adjacent prompts every 5 seconds, which requires three interpolated illustrations. At the same time, to make the video look more coherent, we employ the morphing~\cite{wolberg1998image} technique to generate more transition frames between every two illustrations. Finally, we combine all the illustrations and frames, and add the input music to the audio track to generate an MP4 video.

\section{Evaluation}

\label{sec:tech}
The performance of the \name system and our lyric generation model was analyzed through a comprehensive evaluation, comprising both quantitative experiments and a user study. To begin with, the lyric generation capabilities of our model were compared to those of two established baseline models. Thereafter, a user study was conducted to compare the quality of music videos produced by our model and WZRD, a state-of-the-art music visualization platform~\cite{wzrd}.

\subsection{Quantitative Experiments}

To demonstrate the performance of our lyric generation model, we compared the quality of the lyrics generated by \name and two baseline models from the aspect of creativity, coherence, and lyric-illustration~compatibility. 

\textbf{Baseline models.}  The majority of existing research on Automated Lyrics Generation~(ALG) has focused on generating lyrics from textual contexts or melodies. To the best of our knowledge, there have been limited studies on ALG specifically for music, and even fewer for practical applications such as language learning. In light of this, to assess the efficacy of the proposed system for this task, the following two state-of-the-art baseline models were selected:

\begin{itemize}[leftmargin=20pt,topsep=1pt,itemsep=1px]

\item{\textit{Transformer.}} We built a transformer model, represented as $\mathit{B}_1$,  and trained it using the maximum likelihood estimation, with musical clips serving as the input, and lyrics serving as the output.

\item{\textit{ Music encoder + GPT-2 w/o cross-attention .}} We built another encoder-decoder model~($\mathit{B}_2$), which has a similar structure to our lyric generation but lacks cross-attention between the encoder and decoder. The input music clip is initially processed by the music encoder, and the resulting music representation is concatenated with the embeddings of the previously generated lyrics. This concatenated representation is then input into GPT-2 for the generation of the next lyric line.

\end{itemize}

\textbf{Metrics.} Assessing the quality of lyrics is a complex task due to the diverse requirements involved, particularly when the lyrics are intended for use in language learning. To examine the performance of our proposed system, we propose the utilization of five objective metrics, including novelty, diversity, coherence, BLEU~\cite{papineni2002bleu}, and lyric-illustration compatibility. Considering the lyric lines in the dataset are short sentences, we choose 2-gram-based metrics and 3-gram-based metrics to evaluate the performances of models:

\begin{itemize}[leftmargin=20pt,topsep=1pt,itemsep=1px]

\item{\textit{Novelty.}} The novelty is calculated as the ratio of the number of infrequent n-gram phrases to the total number of n-gram phrases. In accordance with~\cite{xu2017neural,ma2021ai}, we preprocessed the generated lyrics by removing stop words and retaining only the content words. Then, any phrases that did not appear among the most frequent 2000 phrases were considered to be infrequent. 

\item{\textit{Diversity.}} The diversity is the proportion of the distinct n-gram phrases out of the total number of n-gram phrases in the generated lyrics. It has been introduced to measure the diversity of generated text as outlined in~\cite{li2015diversity}.

\item{\textit{Coherence.}} The coherence of the generated lyrics was evaluated through lexical cohesion~\cite{zhao2022discoscore}, which refers to the number of repeated words within the generated text. Specifically, we count the number of repeated words in the lyrics generated for each music and calculate their average to get the final score.

\item{\textit{BLEU.}} The BLEU score~\cite{papineni2002bleu} is calculated by determining the number of overlapping n-grams between the generated lyrics and the reference lyrics. It is a metric used to evaluate the compatibility between generation and reference in machine translation tasks. We employed it to measure the overlaps between generated lyrics and labeled lyrics. 

\item{\textit{Compatibility.}} 
CLIPScore~\cite{hessel2021clipscore} is a measure of the cosine similarity between visual embedding for an image and textual embedding for a caption. We adopt it to evaluate lyric–illustration compatibility. CLIPScore is utilized to judge whether the generated lyrics can be used as prompts to generate suitable illustrations. A higher CLIPScore indicates that the lyrics and illustration match better.

\end{itemize}

\textbf{Data.} To evaluate the performance of our proposed lyric generation model, we randomly selected 100 songs stratified by genres from the test set introduced in section.~\ref{sec:dataset} and generate lyrics using music clips. 

\textbf{Results.} The results of this evaluation are presented in Table~\ref{tab:evaluation}. The results demonstrate that our proposed model outperforms the baseline models in all five metrics. The improvement in the Distinct-2 and Novelty-2 metrics can be attributed to the incorporation of a VAE structure, which enhances the creativity of the generated content by randomly sampling latent vectors in the audio representation space. Furthermore, the pre-trained GPT-2 model provides substantial support for generating diverse text, outperforming the Transformer baseline model. Additionally, our model's relatively high lexical repetition score indicates its ability to generate meaningful and coherent lyrics.

In summary, the results of our evaluation demonstrate the superiority of our proposed lyric generation model compared to the baseline models. The utilization of a VAE structure and the pre-trained GPT-2 model contributes to the improvement in the generation of diverse and creative lyrics, as well as in terms of meaningfulness and coherence. These results provide valuable insights into the potential of our proposed model for generating high-quality lyrics from~musical~inputs.

\begin{figure*}[!htb]
\setlength{\abovecaptionskip}{10pt}
\centering 
\includegraphics[width=0.9\textwidth]{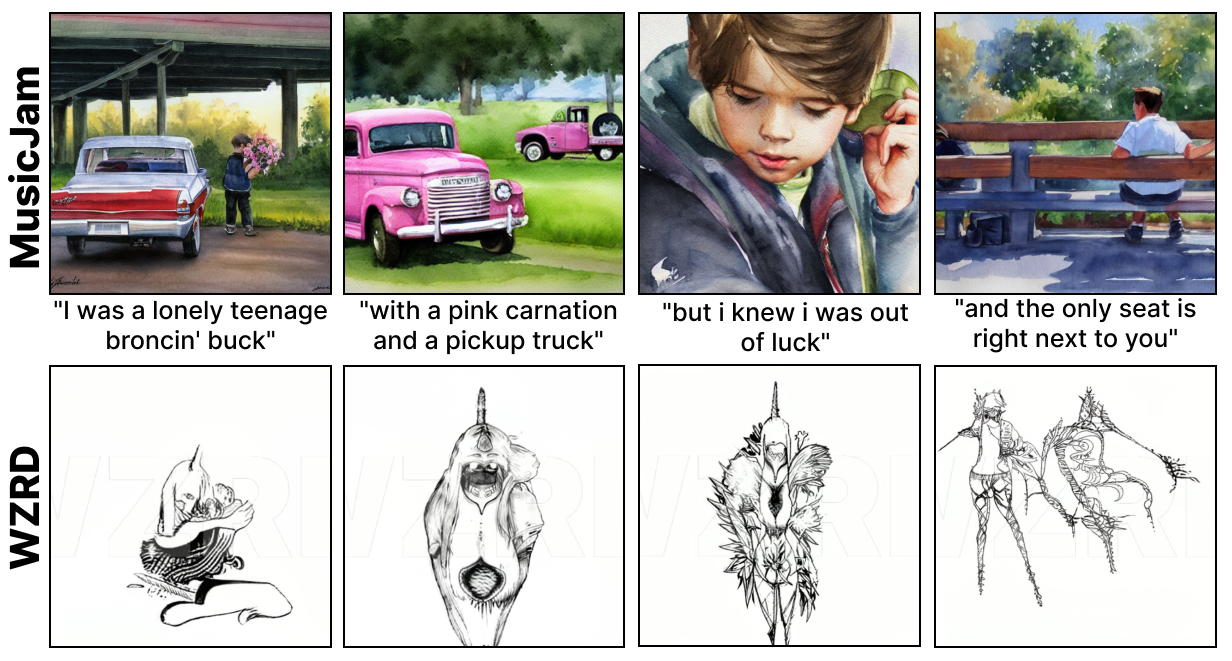}
\vspace{-0.5em}
\caption{Key frames of music videos created by \name and WZRD based on a piece of alternative music, ``Yellow''. The music video created by \name can be found in the following link: \url{https://www.youtube.com/shorts/zyPMSkY98Io}
}
\label{fig:alternative}
\vspace{-0.5em}
\end{figure*}

\begin{figure*}[!htb]
\setlength{\abovecaptionskip}{10pt}
\centering 
\includegraphics[width=0.9\textwidth]{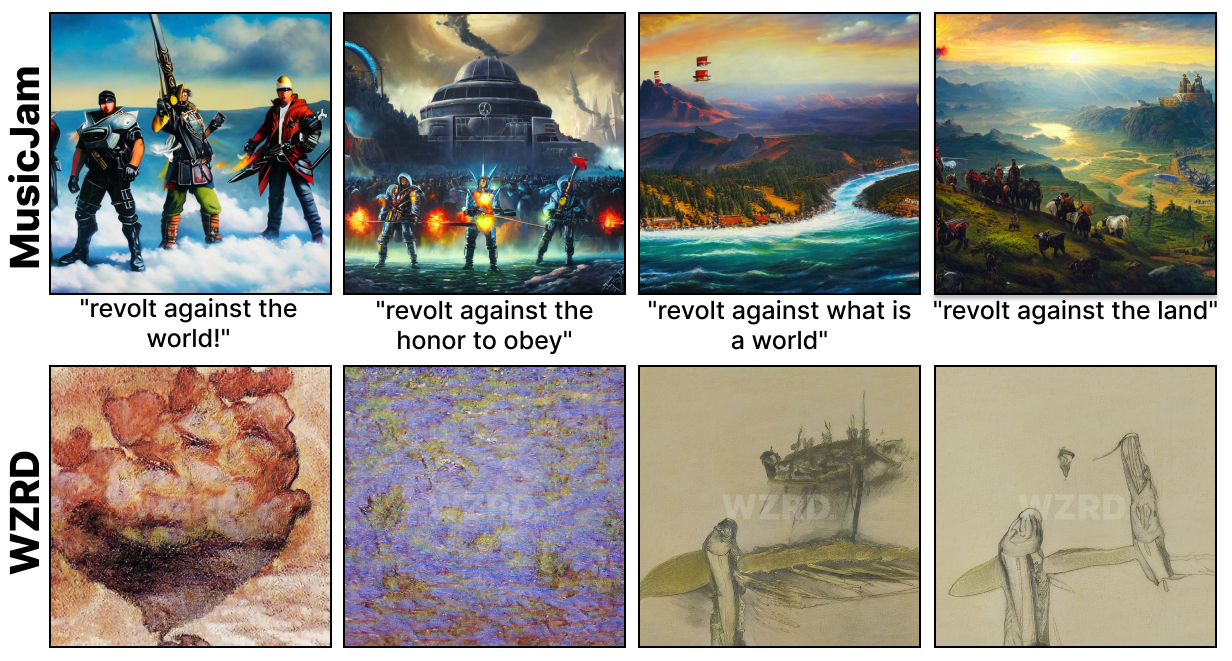}
\vspace{-0.5em}
\caption{Key frames of music videos created by \name and WZRD based on a piece of electronic music, ``Promises''. The music video created by \name can be found in the following link: \url{https://www.youtube.com/shorts/-j615o32Yb8}
}
\label{fig:electronic}
\vspace{-1em}
\end{figure*}

\subsection{User Study}
To assess the quality of the music video generated by the \name system, we conducted a user study for comparison with a state-of-the-art music visualization tool, WZRD~\cite{wzrd}. Given the lack of a precise criterion for subjective evaluation of music video's quality, we integrated subjective metrics from the fields of automatic lyric generation and automatic story scene generation~\cite{ailyric}\cite{storygan}. The quality of the music video was evaluated based on five metrics, including the relevance between the music and videos, the aesthetic appeal of the videos, the meaningfulness of the videos, the coherence of the videos, and the comprehensibility of the videos.

\textbf{Data.}
In our user study, we generated 14 videos, with half being generated by the \name system and half by WZRD. The input music for these videos was sourced from seven different music genres, including pop, alternative, rock, dance, country, electronic, and metal, to accommodate a broad range of common music genres. One song was selected for each genre. In order to ensure participant engagement and avoid misunderstandings, all input music was trimmed to 30 seconds, as a song that is either too long may result in participant boredom or too short may result in participant confusion regarding the music video. Fig.~\ref{fig:alternative} and fig.~\ref{fig:electronic} are two comparison cases used in our study, they are generated by alternative music and electronic music respectively.

\textbf{Procedure.}
For the purpose of this experiment, 40 participants were recruited, comprising 20 designers (16 males and 4 females, with a mean age of 24.15) and 20 non-designers (14 males and 6 females, with a mean age of 23.8). The order of presentation of the 14 videos was randomized and each video was shown to a participant one at a time. Following the viewing of each video, participants were asked to provide ratings on a 5-point Likert scale~(1 =``strongly disagree", 5=``strongly agree") regarding 5 metrics, including the relevance between the music and videos, the coherence of videos, the aesthetic of the videos, the meaningfulness of the videos, and the comprehensibility of the videos. 

\textbf{Hypothesis.}
Prior to conducting the formal user study, a pilot study was conducted using 6 videos to compare the videos generated by the \name system and WZRD, with 10 participants. The design of the pilot study was identical to that of the formal study. Based on the results of the pilot study, the following hypotheses were formulated:

\begin{enumerate}[topsep=1pt,parsep=0pt, itemsep=1pt]
\item[{$H1$}] \name should generate music videos that surpass WZRD in terms of the compatibility between music~and~videos.
\item[{$H2$}] \name is projected to generate music videos that surpass WZRD in terms of aesthetic appeal.
\item[{$H3$}] In terms of meaningfulness, the music videos produced by \name are expected to outperform that of WZRD. 
\item[{$H4$}] \name's generated videos are anticipated to exhibit a higher level of coherence superiority compared to~WZRD.


\item[{$H5$}] The comprehensibility of the music videos generated by \name are expected to surpass that of WZRD. 
\end{enumerate}

\begin{figure}[!tbh]
  \centering
  \includegraphics[width=\linewidth]{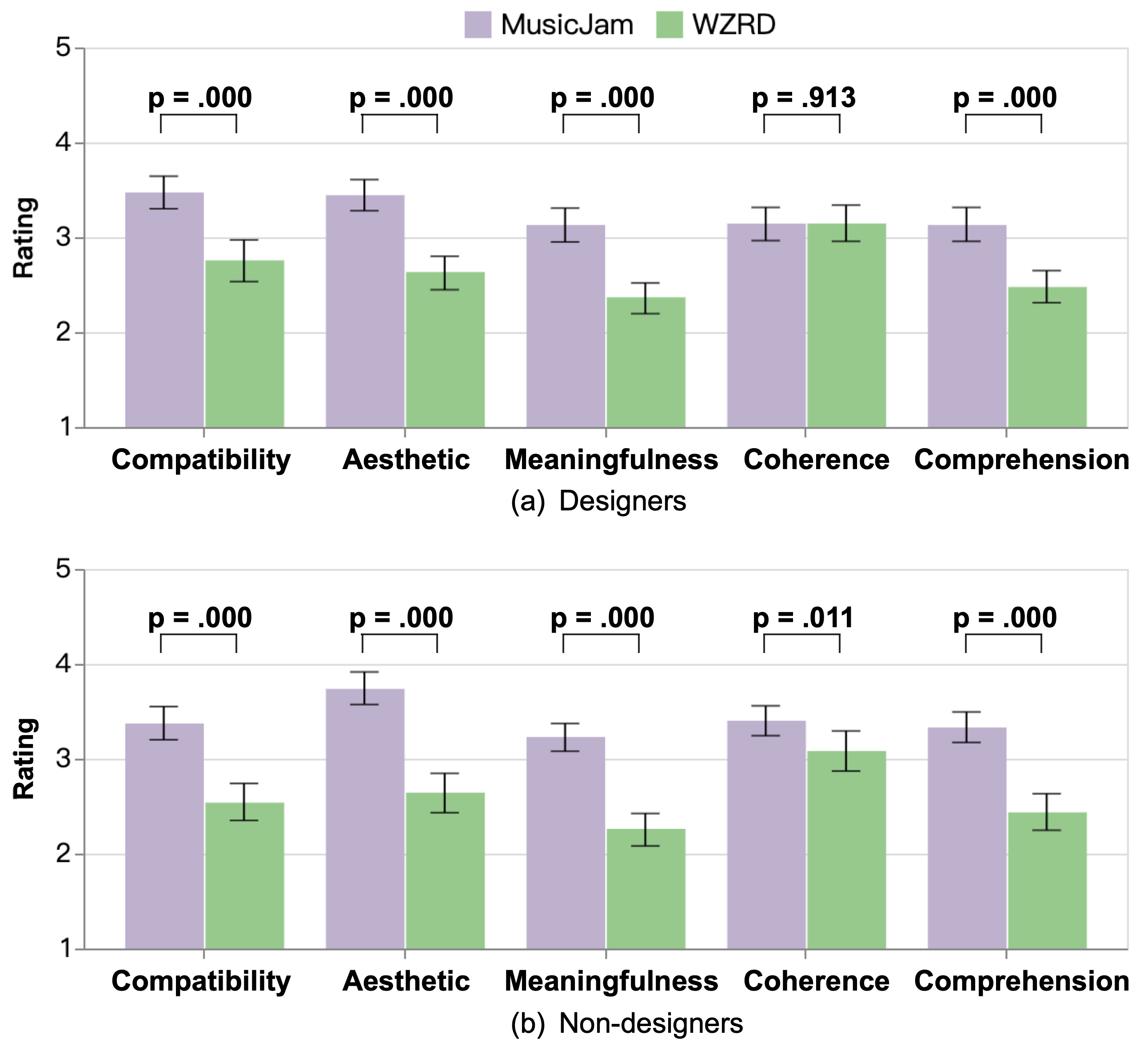}
  \vspace{-1.5em}
  \caption{The ratings of music visualizations from different criteria based on a 5-point Likert scale given by 20 designers and 20 non-designers, where 5 is the best and 1 is the worst.} 
  \label{fig:userstudyresult}
  \vspace{-1em}
\end{figure}

\textbf{Results.}  The study collected 2800 ratings from 40 participants among 14 music videos. In order to gain a deeper insight into how individuals with differing backgrounds perceive the generated video, the ratings were separated into two groups based on the participant's background as either a designer or a non-designer. The results were presented separately in~Fig~.\ref{fig:userstudyresult}.

\textit{Compatibility.} Designers' ratings indicate that \name (M=3.47, SD=1.06)  significantly outperformed WZRD (M=2.76, SD=1.24) in terms of music-videos compatibility based on the t-test~($t(140) = 5.128$, p$<$0.01). Similarly, non-designers' ratings showed that \name (M=3.37, SD=1.13 ) were significantly better than ($t(140) = 5.528$, p$<$0.01) WZRD averaged (M=2.63, SD=1.11). Thus, $H1$~was~accepted. 

\textit{Aesthetic.} The aesthetic of the videos generated by \name~(M=3.44, SD=1.02) was deemed significantly superior to that of WZRD~(M=2.64, SD=1.06) based on the ratings of designers and the results of a t-test~($t(140) = 6.436$, p$<$0.01)). Moreover, non-designer ratings indicated that \name~(M=3.74, SD=1.02) was also significantly better than WZRD~(M=2.69, SD=1.19) according to the results of a t-test ($t(140) = 7.878$, p$<$0.01). Therefore, the hypothesis~$H2$~was~confirmed.

\textit{Meaningfulness.} Based on designer ratings and the results of a t-test~($t(140) = 6.154$, p$<$0.01), the meaningfulness of the videos generated by \name~(M=3.13, SD=1.07) was found to be significantly superior to that of WZRD~(M=2.36, SD=1.01). Additionally, non-designer ratings showed that \name~(M=3.23, SD=0.92) was significantly better than WZRD~(M=2.29, SD=1.00) based on t-test~($t(140) = 8.232$, p$<$0.01). Hence, the hypothesis $H3$ was accepted.

\textit{Coherence.} User ratings from non-designers revealed that the coherence of videos generated by \name~(M=3.40, SD=0.97) significantly exceeded that of WZRD~(M=3.06, SD=1.21) based on the results of a t-test~($t(140) = 2.570$, p$<$0.05). However, designer ratings demonstrated that \name~(M=3.14, SD=1.04) was not superior to WZRD~(M=3.16, SD=1.15) according to t-test ($t(140) =-0.109$, p$=$0.913). Consequently, the hypothesis $H4$ was rejected.

\textit{Comprehensibility.} The comprehensibility of the videos generated by \name~(M=3.13, SD=1.09) was deemed significantly superior to that of WZRD~(M=2.47, SD=1.04) based on the ratings of designers and the results of a t-test ($t(140) = 5.152$, p$<$0.01). Moreover, non-designer ratings indicated that \name~(M=3.33, SD=1.00) was also significantly better than WZRD~(M=2.47, SD=1.08) according to the results of a t-test ($t(140) = 6.883$, p$<$0.01). As a result, the hypothesis $H5$ was confirmed.

The above analysis demonstrates that the ratings provided by designers and non-designers were roughly comparable. However, the ratings regarding coherence were disparate. Two potential explanations for this result are proposed. Firstly, in order to enhance the coherence of the videos, WZRD incorporated transitions between frames while \name utilized interpolation between frames. This resulted in the participants perceiving minimal disparities in terms of coherence while viewing the videos. Secondly, the participants, who are designers, possessed professional technical expertise and thus may have held a higher standard for evaluating the videos.


\section{LIMITATIONS AND FUTURE WORK}
\label{sec:limitations}

Despite the positive evaluation results indicating \name is promising to visualize music via generated narrative illustrations, the system still has several limitations that were found during the design and implementation process. We hope to point out several potential future research directions by highlighting these limitations. 

\underline{\textit{Improving Video Coherence.}}
The music visualizations created by \name do not exhibit a high level of coherence as evaluated in the user study. We suspect that is due to the number of interpolated illustrations being few, resulting in less smooth transitions between illustrations. The generation time increases in proportion to the number of interpolations. Thereby, to ensure a better user experience, we reduced the number of interpolations. If the illustration generation process can be accelerated in the future, we can interpolate more illustrations to provide users with a more coherent visual experience.


\underline{\textit{Improving Prompt Generation Quality.}}
In this work, we primarily focused on generating lyrics that match the music but ignored the lyric-illustration compatibility. The training process of the lyrics generation model did not consider how to generate appropriate lyrics that can produce high-quality illustrations. As a result, some prompts built from the lyrics lead to unattractive or meaningless generated illustrations. To address this issue, future work should take lyric-illustration compatibility into consideration when designing the model and~training~objectives. 



\underline{\textit{Augmenting Training set.}}
The lyric generation model was trained on music from various genres. However, the number of music in each genre is unbalanced where pop music makes up the majority of the training data. This issue results in the tendency to generate lyrics with common content, such as love stories. To eliminate this problem and generate more diverse lyrics, a more diverse and balanced training set is desired to be collected further and used to train the model.

\underline{\textit{Accelerating Generation Process.}}
The current system design and implementation have some performance bottlenecks. In particular, it usually takes a considerable time to generate a one-minute music video. This is due to the high computational cost required by stable diffusion. To handle this situation, we should either upgrade the computing device or utilize faster diffusion models to accelerate the~generation~process.

\section{CONCLUSION}
\label{sec:conclusion}
In this paper, we present MusicJam, a music visualization system designed for generating narrative illustrations to represent the insights of invisible music. The system incorporates a novel generative model to generate lyrics given the input music and then transforms the lyrics into a sequence of coherent illustrations using an interpolation method. The proposed technique was evaluated via quantitative experiments and a controlled user study. The evaluation showed that the music videos generated by MusicJam were rated higher regarding quality compared with WZRD, which demonstrated the effectiveness of our method. We believe our technique can contribute to broadening the music visualization field, helping users better understand the insights of music and enjoy an immersive listening experience. 

\acknowledgments{
Nan Cao is the corresponding author. This work was supported by NSFC 62072338, NSFC 62061136003. }

\bibliographystyle{abbrv-doi-hyperref}

\bibliography{template}


\end{document}